\documentclass[aps,prl,preprint,superscriptaddress,showpacs]{revtex4}

\usepackage[dvips]{graphicx,color}% Include figure files
\usepackage{dcolumn}% Align table columns on decimal point
\usepackage{bm}% bold math

\begin{document}

\preprint{}

\title{
Collinear-to-Spiral Spin Transformation without Changing Modulation Wavelength upon Ferroelectric Transition in Tb$_{1-x}$Dy$_{x}$MnO$_{3}$
}

\author{T. Arima}
\affiliation{Institute of Multidisciplinary Research for Advanced Materials,
Tohoku University, Sendai 980-8577, Japan}
\affiliation{Spin Superstructure Project, ERATO, Japan Science and Technology Agency, 
AIST Tsukuba Central 4, Tsukuba 305-8562, Japan
} 
\author{A. Tokunaga}
\affiliation{Institute of Materials Science, University of Tsukuba,
Tsukuba 305-8573, Japan}

\author{T. Goto}
\affiliation{Department of Applied Physics, University of Tokyo,
Tokyo 113-8656, Japan
}

\author{H. Kimura}
\affiliation{Institute of Multidisciplinary Research for Advanced Materials,
Tohoku University, Sendai 980-8577, Japan}
\author{Y. Noda}
\affiliation{Institute of Multidisciplinary Research for Advanced Materials,
Tohoku University, Sendai 980-8577, Japan}

\author{Y. Tokura}
\affiliation{Spin Superstructure Project, ERATO, Japan Science and Technology Agency, 
AIST Tsukuba Central 4, Tsukuba 305-8562, Japan
}
\affiliation{Department of Applied Physics, University of Tokyo,
Tokyo 113-8656, Japan
}

%%%%%%%%%%%%%%%%% END OF PREAMBLE %%%%%%%%%%%%%%%%

\date{\today}

\begin{abstract}
Lattice modulation and magnetic structures in magnetoelectric compounds Tb$_{1-x}$Dy$_{x}$MnO$_3$ have been studied around the ferroelectric (FE) Curie temperature $T_C$ by x-ray and neutron diffraction.  
Temperature-independent modulation vectors through $T_C$ are observed for the compounds with $0.50\le x \le 0.68$.  This indicates that ferroelectricity with a polarization ($P$) along the $c$ axis in the $R$MnO$_3$ series cannot be ascribed to such an incommensurate-commensurate transition of an antiferromagnetic order as was previously anticipated.  
Neutron diffraction study of a single crystal with $x=0.59$ shows that the FE transition is accompanied by the transformation of the Mn-spin alignment from sinusoidal (collinear) antiferromagnetism into a transverse spiral structure.  The observed spiral structure below $T_C$ is expected to produce $P$ along the $c$ axis with the `inverse' Dzialoshinski-Moriya interaction, which is consistent with the observation.   

\end{abstract}


\maketitle
%%%%%%%%%%%%%%%%%%%%%%%%%%%%%%

%%% Introduction:

Research on coupling between ferroelectric (FE) and magnetic orders has revived after a large magnetoelectric (ME) response was reported in several compounds.\cite{Fiebig_review}  
In particular, gigantic ME effects in rare-earth manganites $R$MnO$_3$  ($R=$Gd, Tb, or Dy) with orthorhombically distorted perovskite structure have stimulated considerable interest since Kimura {\it et al.} reported that an electric polarization $P$ can be flopped by applying a magnetic field in TbMnO$_3$.\cite{Kimura_nature}  
Evidence has been increasing that this unique ME effect is inherently connected to the Mn-spin ordering structure.\cite{Kimura_nature,Kimura_PRB1,Goto,Kimura_PRB2}  
TbMnO$_3$ and DyMnO$_3$ undergo an antiferromagnetic (AFM) transition with a temperature-dependent modulation vector (0 $q_{\rm Mn}$ 1) around 40 K.  Hereafter, we use the {\it Pbnm} orthorhombic setting.  With further cooling, the $q_{\rm Mn}$ value appears to be locked, and simultaneously, FE polarization appears along the $c$ axis.
GdMnO$_3$, whose orthorhombic distortion is less than that of TbMnO$_3$, shows a magnetic-field-induced FE state with $\makebox{\boldmath $P$}\parallel a$.\cite{Popov,Noda,Kimura_PRB2}  
Synchrotron x-ray diffraction measurements of GdMnO$_3$ and TbMnO$_3$ in magnetic fields revealed that the modulation wavevector changes with the magnetic-field-induced electric transitions.\cite{Arima} 

\begin{figure}
\includegraphics*[width=7.5cm]{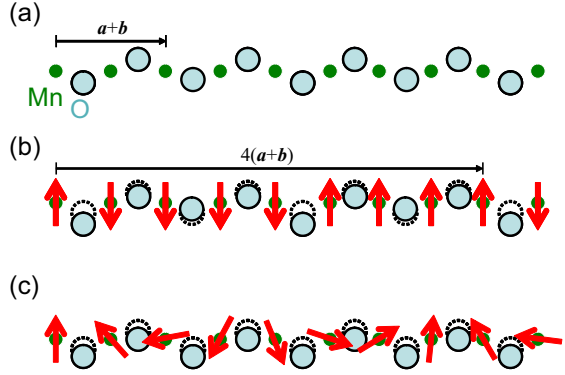}
\caption{
(Color online) 
Two possible mechanisms of spin-order-driven ferroelectricity in orthorhombic $R$MnO$_3$.
(a) 
A zigzag Mn--O chain along $\left<110 \right>$, {\it i.e.}, in the basal plane (001).  
All the bonds bend with the GdFeO$_3$-type distortion above the N\'eel temperature $T_N$.   Small and large balls represent Mn and O ions, respectively.  
(b)
Ferroelectricity induced by antiferromagnetic ordering of Ising-type Mn spins with $q=1/4$ and the resultant exchange striction. 
Oxygen ions located between up and down spins are displaced in the same direction.  
A similar situation can be expected when the modulation vector of the Mn-spin moments (not necessarily collinear) is $m/n$, with odd $m$ and even $n$.  
(c)  
All the oxygen ions are displaced in the same direction with a transverse spiral alignment of the Mn-spin moments.  
}
\label{mechanisms}
\end{figure}

There are two possible mechanisms of spin-order-driven ferroelectricity (see Fig.~\ref{mechanisms}).  
It was suggested that the ferroelectricity in $R$MnO$_3$ can be attributed to the {\it symmetric} exchange striction at the earlier stage, due to the locking of $q_{\rm Mn}$ at the FE Curie temperature $T_C$.
The superexchange interaction via an anion $X$ between two magnetic cations ($M$) can be modified with an $M$--$X$--$M$ angle.\cite{GK}  Conversely, an $M$--$X$--$M$ bond angle should be more or less affected by the spin arrangement.  
In perovskite rare-earth manganese oxides $R$MnO$_3$, Mn--O--Mn bonds are not straight due to rotation and tilt of the MnO$_6$ octahedra.  
Therefore, the Mn--O--Mn--O-- bonding along $\left< 110\right>$ forms a zigzag chain, as schematically shown in Fig.~\ref{mechanisms}(a).  
Below the AFM transition temperature $T_N$, each Mn--O--Mn angle may be further influenced by the alignment of Mn moments.  
The modification of the oxygen positions with an AFM spin arrangement is demonstrated in an exaggerated manner in Fig.~\ref{mechanisms}(b), where the oxygen ions are represented first by dotted circles and then by solid ones.   
Ferromagnetic superexchange coupling between two neighboring Mn ions in the basal plane becomes weaker as the Mn--O--Mn bond angle decreases.\cite{Kimura_PRB1,GK}  
The inverse interaction should decrease the angle of the Mn--O--Mn bonds between antiferromagnetically aligned neighboring Mn spins, thereby resulting in a lattice modulation with a propagation vector $q_L=2q_{\rm Mn}$.  
In the original zigzag chain, the oxygen ions between up and down spins move in the same direction when the magnetic modulation vector is represented as $1/n$, where $n$ is an even integer.  
More generally, a periodic modulation of $ \makebox{\boldmath $S_i$}\cdot \makebox{\boldmath $S_j$} $ with a modulation vector of $m/n$ can induce ferroelectricity in the chain, when $m$ and $n$ are odd and even integers, respectively.  

% inverse DM
As an alternative mechanism of the spin-order-driven ferroelectricity, it has recently been pointed out that transverse spiral spin ordering can induce ferroelectricity regardless of the commensurability between spin modulation and the lattice, as shown in Fig.~\ref{mechanisms}(c).\cite{Katsura,Sergienko}  
This mechanism is closely related to the {\it antisymmetric} superexchange or the so-called Dzialoshinski-Moriya (DM) interaction.\cite{DM}  A bending $M_i$--$X$--$M_j$ bond lacking an inversion center favors a canted spin arrangement.  This antisymmetric superexchange term is linear to the spin-orbit coupling term, and is expressed as $\makebox{\boldmath $D$}\cdot \left(\makebox{\boldmath $S_i$}\times\makebox{\boldmath $S_j$}\right)$, where {\boldmath $D$} is a constant vector.  
By considering the inverse effect, the stable position of the $X$ ion between two magnetic $M$ ions can be modified using the vector product of the two magnetic moments, $\makebox{\boldmath $S_i$}\times\makebox{\boldmath $S_j$}$.  Since the vector product of any two neighboring $M$ moments is constant in case of transverse spiral ordering, all the $X$ ions are displaced in the same direction.

Recent neutron diffraction studies have demonstrated a change in the magnetic structure of TbMnO$_3$ at $T_C$.\cite{Kajimoto,Kenzelmann}   
Kenzelmann {\it et al.} discussed the magnetic symmetry by means of a model fitting of $1q$ components of magnetic reflections, which indicated a sinusoidal (collinear) and spiral (non-collinear) spin ordering above and below $T_C$, respectively.
From their precise measurements, it was also found that the modulation vector of antiferromagnetism is nearly locked at temperature greater than $T_C$ by a few K.\cite{Kenzelmann}   
These results are inconsistent with the exchange striction mechanism.  
However, the spiral spin structure could not be evidenced clearly because the higher Fourier component was neglected.  
%Therefore, it is still not clear whether the incommensurate-commensurate transition in spin ordering causes the FE lattice modulation, which in turn modifies the spin structure through DM interaction, or otherwise, the spiral spin ordering itself is the origin of electric polarization.    
%

This paper reports that it is the spiral spin ordering and not the locking of $q_{\rm Mn}$ that plays a dominant role in producing ferroelectricity in $R$MnO$_3$.   
Up to $T_N$ and even above $T_C$, solid solution of TbMnO$_3$ and DyMnO$_3$ in an appropriate ratio shows long-wavelength antiferromagnetism with a temperature-independent modulation vector.  
A neutron diffraction study on Tb$_{0.41}$Dy$_{0.59}$MnO$_3$ with $q_{\rm Mn}\approx 1/3$ enables us to unambiguously determine the spin structure without a model fitting.  
The results clearly indicate a spiral magnetic ordering below $T_C$, which changes from a collinear structure with an identical $q_{\rm Mn}$ ($\approx 1/3$). 

%%%%%%%%%%%%%%%%%%%%%%%%%%% figure 1 %%%%%%%%%%%%%%%%%%%%%%%%%%%%%%%%
%%%%%%%%%%%%%%%%%%%%%%%%%%% figure 1 %%%%%%%%%%%%%%%%%%%%%%%%%%%%%%%%

%%% Experimental:
Single crystals of Tb$_{1-x}$Dy$_x$MnO$_3$ were grown by a floating zone method.\cite{Kimura_PRB1}   
Off-resonant single-crystal x-ray diffraction measurements were performed on Beamline 4C at Photon Factory, JAPAN.  
A 13-keV monochromatic x ray was focused on a single crystal in a closed-cycle He refrigerator, which was mounted on a four-circle diffractometer.  
%

%%% Locking of (Tb,Dy)MnO$_3$ %%%

\begin{figure}
\includegraphics*[width=8cm]{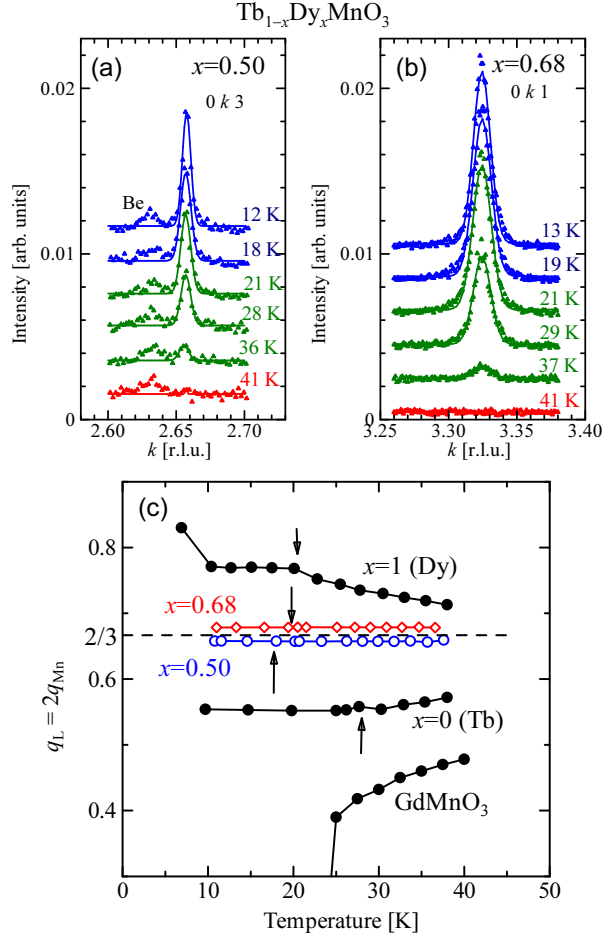}
\caption{
(Color online)
(a) and (b) X-ray diffraction profiles of (0 k 1) scan in Tb$_{1-x}$Dy$_x$MnO$_3$ with $x=0.50$ and $0.68$ at various temperatures.  
A peak at $k\approx 2.63$ for $x=0.50$ appears even above $T_N$ and can be ascribed to x-ray diffraction from a Beryllium window.  
(c) Lattice modulation wavenumber $q_L$ in some Tb$_{1-x}$Dy$_x$MnO$_3$ compounds as a function of temperature.     
Here, $q_L=2q_{\rm Mn}$ (see text).  
Arrows indicate the ferroelectric Curie temperatures.  
}
\label{qT}
\end{figure}

First, we show that the lattice modulation vector $q_L$ can be locked even above $T_C$ in some solid solutions of TbMnO$_3$ and DyMnO$_3$.   
Figures \ref{qT}(a) and (b) show the evolution of a superlattice x-ray diffraction peak with temperature in Tb$_{1-x}$Dy$_x$MnO$_3$ with $x=0.50$ and 0.68.  A superlattice reflection is observed at (0 2.66 3) and (0 3.32 1) for $x=0.50$ and 0.68, respectively, and the $q_L$ value can be regarded as 0.66 and 0.68, respectively.  
The position of the superlattice reflection is not dependent on the temperature for both the compounds.  
In Figure \ref{qT}(c), the $q_L$ values for some compounds are plotted against temperature.  The $T_C$ values are also shown by arrows in the panel.  
Here, it should be noted that the $q_L$ value should be equal to twice the magnetic modulation vector of the Mn ions, $q_{\rm Mn}$.\cite{Kimura_PRB1}  Therefore, we can conclude that the nearly commensurate $q_{\rm Mn}$ values ($\sim 1/3$) of the compounds with $x=0.50$ and 0.68 are independent of temperatures below $T_N$ within the resolution limit, in contrast to TbMnO$_3$ and DyMnO$_3$.  
This result contradicts the simple scenario in which the commensurate locking of the AFM wavevector causes an FE transition in orthorhombic $R$MnO$_3$.  

%%%%%%%%%%%%%%%%%%%%%%%%%%% figure 2 %%%%%%%%%%%%%%%%%%%%%%%%%%%%%%%%
%%%%%%%%%%%%%%%%%%%%%%%%%%% figure 2 %%%%%%%%%%%%%%%%%%%%%%%%%%%%%%%%

%%%%%%%%%%%%%%%%%%%%%%%%%%% magnetic structure %%%%%%%%%%%%%%%%%%%%%%%%%%%%%

Another possible origin of ferroelectricity in the orthorhombic $R$MnO$_3$ system is a change in the spin structure as aforementioned.  
Hence, the determination of the magnetic structure in the Tb$_{1-x}$Dy$_x$MnO$_3$ compound provides us important information for discussing a mechanism of ferroelectricity in the system.  
It is very difficult to carry out a model-free analysis of a magnetic structure with an incommensurate wavenumber because there are infinite independent magnetic sites in an incommensurate antiferromagnet.  
Therefore, we selected a compound with $x=0.59$, where the $q_{\rm Mn}$ value can be regarded as 1/3, judging from the $q_{\rm Mn}$ values corresponding to  $x=0.50$ and 0.68, as shown in Fig.~\ref{qT}.  
The measurements of the position of the magnetic peaks showed that the $q_{\rm Mn}$ value of the compound with $x=0.59$ was also independent of temperature (not shown in with a figure) within the resolution limit of the present neutron diffraction measurement.  
$T_N$ and $T_C$ of the compound are 41 K and 22 K, respectively.  
The AFM ordering of rare-earth moments occurs at 7 K.  

\begin{figure}
\includegraphics*[width=7.5cm]{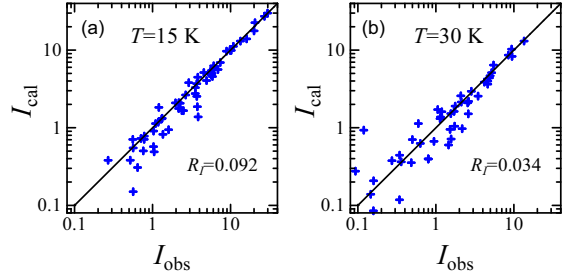}
\caption{
(Color online) Comparison between observed and calculated intensities of magnetic reflections at (a) 15 K and (b) 30 K.   
The calculation was based on the magnetic structures shown in Fig.~4.
}
\label{obscal}
\end{figure}

Single-crystal neutron diffraction measurements were carried out on a four-circle neutron diffractometer (FONDER) installed in the guidehall of JRR-3M at the JAPAN Atomic Energy Research Institute, Japan.\cite{FONDER}  
The FONDER diffractometer was operated with a wavelength $\lambda=1.24$ \AA\ that was obtained by using a Ge (311) monochromator.  
A single crystal with a diameter of $\sim 2.5$ mm and a length of $\sim 7$ mm was used for the neutron measurements.  
Integrated intensities were collected for 176 nuclear reflections at 15 K and 150 and 67 magnetic reflections at 15 K and 30 K, respectively.  
(At 30 K, we did not perform measurements for high-angle magnetic reflections where no peak was discernible at 15 K.)  
%Typical measurement time was 10 minutes per reflection.  
The absorption was corrected by using DABEX software.
The atomic coordination and isotropic thermal parameter of each atom as well as the scaling factor and extinction were determined from the intensities of the nuclear reflections by using a least-square program.   
Here, we neglected the superlattice modulation of the atomic positions because it was too weak to be detected by neutron diffraction.  
The magnetic moments were calculated from the intensity data of the magnetic reflections, including the magnetic contribution at fundamental reflections, by using a Levenberg-Marquardt-type least-square method.  
The magnetic form factors of Mn$^{3+}$, Tb$^{3+}$, and Dy$^{3+}$ were assumed to be isotropic.\cite{international}  
At 30 K, the fitting parameters were three components at each of the twelve Mn ($m_{i,a}$, $m_{i,b}$, and $m_{i,c}$).  At 15 K, the magnetic moments at rare-earth sites were also considered to fit the data.  
Many trials of gradient descent algorithms with different initial guesses verified that the obtained result was one of the best minima.  
Although there are several solutions with almost similar residuals, they are essentially identical: they can transform into each other with some symmetry operation.  
Figure \ref{obscal} presents the comparison between the observed ($I_{\rm obs}$) and calculated ($I_{\rm cal}$) intensities of the magnetic reflections.  
The reliability factors based on integrated intensities, $R_I$, are 0.092 and 0.034 for 15 K and 30 K, respectively.  
The average deviation, defined as
$%\displaystyle
(1/N)\sum {\left(I_{\rm cal}-I_{\rm obs}\right)^2
/ \sigma_I^2},
$
where $N$ denotes the data number, is satisfactorily small for both temperatures (0.7 and 1.1 for 15 K and 30 K, respectively).

%%%%%%%%%%%%%%%%%%%%%%%%%%% figure 3 %%%%%%%%%%%%%%%%%%%%%%%%%%%%%%%%
%%%%%%%%%%%%%%%%%%%%%%%%%%% figure 3 %%%%%%%%%%%%%%%%%%%%%%%%%%%%%%%%

%%%%%%%%%%%%%%%%%%%%%%%%%%% figure 4 %%%%%%%%%%%%%%%%%%%%%%%%%%%%%%%%
\begin{figure}
\includegraphics*[width=8.5cm]{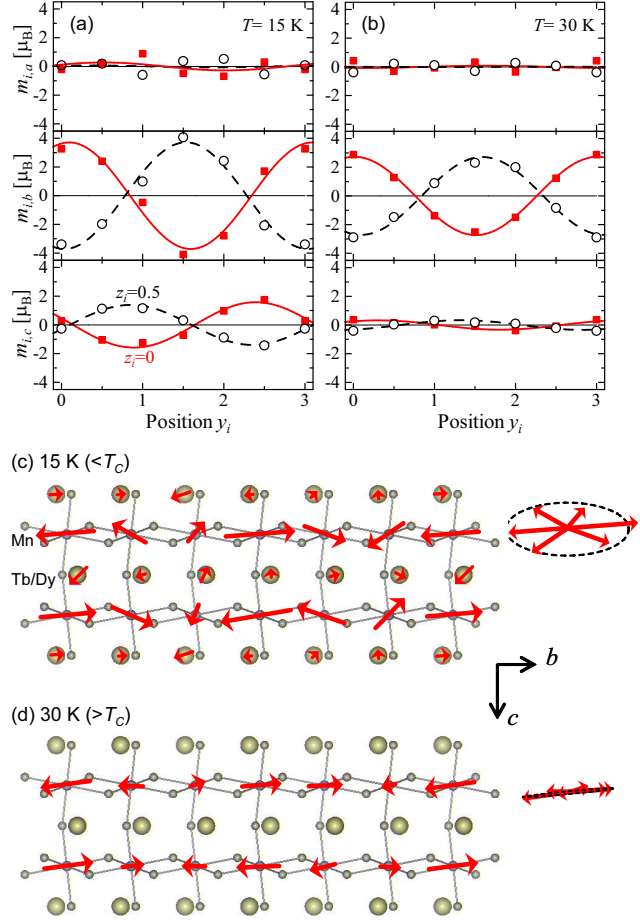}
\caption{
(Color online)
(a) and (b) Alignment of Mn-spin moments at 15 K and 30 K obtained from least square fitting of the intensity of the magnetic reflections.  Note that there are two basal Mn--O planes in a unit cell with $z_i=0$ and 0.5.  Lines show the best  fits with sinusoidal functions.   
(c) and (d) Schematic drawing of the magnetic moments projected onto the (100) plane at 15 K and 30 K.  Right panels show the trajectories of the Mn spins.    
}
\label{spins}
\end{figure}
%%%%%%%%%%%%%%%%%%%%%%%%%%% figure 4 %%%%%%%%%%%%%%%%%%%%%%%%%%%%%%%%

The obtained magnetic structures at 15 K ($< T_C$) and 30 K ($> T_C$) are shown in Fig.~\ref{spins}.  
The $a$ component of any Mn spin is smaller than 1 $\mu_{\rm B}$ at both the temperatures.  
The $b$ components of the Mn spins are sinusoidally modulated along the $b$ axis with an amplitude of 3.7 and 2.7 $\mu_{\rm B}$ at 15 K and 30 K, respectively.  The clearest difference is observed in the modulation of the $c$ components.  
The amplitude of sinusoidal modulation is $1.4\sim 1.6$  $\mu_{\rm B}$ at 15 K, while it is less than 0.4 $\mu_{\rm B}$ at 30 K.  
Below $T_C$, the phase of the modulation of the $c$ components is shifted by $0.45\pi$--$0.47\pi$ from that of the $b$ components.  Note that the phase shift is $\pm\pi/2$ in the case of an ideal spiral structure.
This transverse-spiral alignment of spins breaks mirror reflection symmetry normal to the $c$ plane, in accordance with the $\makebox{\boldmath $P$} \parallel c$ state below $T_C$.  
Further, it should be noted that the trajectory of the Mn spins is not a circle but an ellipsoid, as shown in the inset of Fig.~\ref{spins}(c).  
The value of $\makebox{\boldmath $S_i$}\times \makebox{\boldmath $S_{i+1}$}$ and the displacement of oxygen ions due to the inverse DM interaction should also be modified with $2q_{\rm Mn}$.  
This is consistent with the observed relation $q_L=2q_{\rm Mn}$.   
In contrast to the spin structure at 15 K, the difference in phase of the spin modulation between the $b$ and $c$ components becomes considerably smaller at 30 K.  The calculated difference in phase is $+0.20 \pi$ and $-0.17 \pi$ for sites with $z=0$ and $z=0.5$, respectively.  
The spin moments for Mn and Tb/Dy above and below $T_C$ projected onto the (100) plane are schematically drawn in Figs.~\ref{spins}(c) and (d), respectively.  
The induced moments at the rare-earth sites are approximately 1 $\mu_{\rm B}$ at 15 K.
These figures clearly indicate that the FE transition is accompanied by a spin-state transformation from the sinusoidal collinear to transverse spiral type.  
%
%According to Katsura {\it et al.}, the $P$ value induced by transverse spiral spin order in a $t_{2g}$ system is roughly estimated by $10^4 \times (t_{pd}/\Delta_{pd})^3$ $\mu$C/m$^2$.\cite{Katsura}  Here, $t_{pd}$ and $\Delta_{pd}$ are the transfer integral and energy difference between O $2p$ and transition-metal $3d$, respectively.  Since $(t_{pd}/\Delta_{pd})^3\approx 10^{-1}$, the observed $P$ value of $\approx 6\times 10^2$ $\mu$C/m$^2$ in Tb$_{0.41}$Dy$_{0.59}$MnO$_3$ at 15 K could be well explained by their spin-supercurrent model, neglecting the difference in $3d$ orbital character between the model and the real materials.  

In conclusion, synchrotron x-ray diffraction studies of the (Tb,Dy)MnO$_3$ system excludes the scenario in which an incommensurate-to-commensurate locking of the modulation wavevector of an antiferromagnetic Mn-spin ordering would induce ferroelectricity with $\makebox{\boldmath $P$} \parallel c$.  
A model-free analysis of the spin structure of system with $q_{\rm Mn}\sim 1/3$ indicates that the ferroelectric transition results from the sinusoidal-to-spiral spin-structure transition.  
The close relation between non-collinear spin ordering and electric polarization will result in new methods for discovering new magnetoelectric materials.  

%% Acknowledgments %%
%\begin{acknowledgments}
The authors thank T. Kimura and N. Nagaosa for helpful discussions.  
The assistance provided by Y. Watanabe, S. Nishiyama, and H. Sawa in the synchrotron x-ray measurements is gratefully acknowledged.  
The neutron measurements were performed under PACS No.~5694 of ISSP, University of Tokyo.  
This work was partly supported by Grants-In-Aid for Scientific Research from the MEXT, Japan.  
%\end{acknowledgments}

\end{document}